\input amstex
\documentstyle{amsppt} 
\pagewidth{135mm} 
\pageheight{180mm} 
\NoRunningHeads 
\nologo 
\loadbold
\topmatter
\title The Geometry of Spacetime with superluminal phenomena\endtitle
\author T. Matolcsi$^1$  
\\  
and  
\\ 
W. A. Rodrigues Jr.$^2$  
\endauthor 
\affil 
$^1$Department of Applied Analysis \\  
E\"otvos Lor\'and University \\ 
Budapest, Hungary  
\\ 
$^2$Instituto de Matem\'atica, Estat\'{\i}stica e 
Computa\c{c}\~ao Cient\'{\i}fica \\ Universidade Estadual de Campinas \\ 
CP 6065, CEP 13081--970, Campinas,
SP, Brasil 
\endaffil  
\abstract   
Recent theoretical results show the existence of arbitrary speeds
($0\leq v <\infty$) solutions of all relativistic wave equations.  Some
recent experiments confirm the results for sound waves.  The question
arises naturally: What is the appropriate geometry of spacetime to
describe superluminal phenomena? In this paper we present a spacetime
model that incorporates the valid results of Relativity Theory and yet
describes coherently superluminal phenomena without paradoxes.
\endabstract 
\endtopmatter 

\def\M{\bold M}\def\E{\bold E}\def\I{\bold I}\def\ta{\boldsymbol\tau}
\def\D{\bold D}\def\u{\boldkey u}\def\e{\boldkey e}\def\q{\boldkey q}
\def\t{\boldkey t}\def\l{\kappa}\def\h{\boldkey h}
\def\v{\boldkey v}\def\1{\boldkey 1}
\def\s{\boldkey s}\def\x{\boldkey x}\def\w{\boldkey w}
\def\r{\boldkey r}\def\y{\boldkey y}

\pagebreak 

\subhead 1. INTRODUCTION \endsubhead

Recently it was found that wave equations admit solutions which describe 
waves propagating slower or faster than the velocity appearing in the 
equation in question ([1--6]), and there are experiments proving the 
existence of such waves in the case of sound (supersonic waves)~[7]. 
As a particular case, the Maxwell equations, too, admit subluminal and 
superluminal wave solutions with arbitrary speed. If such superluminal
 phenomena exist in Nature then we must reapprise our notions about  
synchronization, future, past etc. The need of  synchronizations different 
from the standard one emerged from the point of view of tachyons [8--10]
but the possibility of superluminal phenomena offers another way. 

Now we try to establish the structure of spacetime deriving from the
existence of superluminal phenomena. Our treatment is somewhat different 
from the usual approaches based on coordinates and transformation rules. 
The mathematical structure of general relativity based on global analysis 
on manifolds taught us that instead of relative quantities (coordinates, 
electric and magnetic field etc) and their transformation rules,  we have 
to work with absolute  quantities (spacetime, electromagnetic field etc.) 
and their splitting according to observers (in time and space, in electric 
and magnetic field etc.). There are such treatments of non-relativistic 
spacetime and special relativistic spacetime [11--14] which show very well 
that the point of view of absolute objects admits a clear and simple 
presentation and excludes the possibility of misunderstanding because 
the rigorous mathematical structure rules out intuitive notions. In the 
usual approach observers (reference frames), coordinate systems are intuitive 
notions and one uses "natural" tacit assumptions. A good example for a 
misleading tacit assumption is that "if he moves at velocity  $v$ relative 
to you then you moves at velocity $-v$ relative to him". It turns out, however, 
that this does not hold in the special relativistic spacetime
(see ref.~[12], \S~II.4.2); the velocity addition paradox~[15] is the consequence
of this incorrect tacit assumption. 

It is often emphasized that coordinates are labels, not physical entities.
On the contrary, splitting of spacetime, spacetime vectors, tensors etc.
has a physical meaning: the split quantities describe how an observer
perceives absolute objects (the splitting of spacetime by an observer gives
the time and the space of the observer, the splitting of electromagnetic
field gives the observed electric and magnetic field etc.) 

\subhead 2. PRELIMINARIES \endsubhead 

We intend to define a mathematical model of spacetime based on experimental 
facts and theoretical assumptions. The basic experimental facts regarding 
inertial reference frames (observers) are the following. Observers
in different inertial reference frames measure time by the ``same'' 
clocks and synchronization process and measure space by the ``same'' rods.
The term ``same'' means a prescription such as: time is measured by the
oscillations of a  cubic crystal consisting of a given number of molecules
of a given material (e.g. quartz), 
and space is measure by a sideline of that crystal. Then it is found that

\smallskip
\noindent 1. Time has
\nopagebreak 
\item \llap{a)} a one dimensional affine structure (time translations are
meaningful),

\item \llap{b)} an orientation (past and future are distinct);

\smallskip
\noindent 2. Space has

\item \llap{a)} a three dimensional affine structure (space translations are
meaningful),

\item \llap{b)} an orientation (right and left are distinct by the decay of
K~mesons),

\item \llap{c)} a Euclidean structure (distances and angles are meaningful).

\smallskip
\noindent 3. The affine structures of time and space are related to each other by 
uniform motions on straight lines; uniform motion relative to an inertial 
reference frame seems a uniform motion  relative to another inertial
reference frame also.

\smallskip
\noindent 4. Time and space in a given inertial reference frame
are related to time and space of other reference frames
(transformation rules).

\smallskip
Then 1, 2a, 2b and 3 suggest that spacetime is a four dimensional oriented 
affine space.

The other structures are deduced from 3 and 4; the different spacetime models 
come from the different meaning of Euclidean structures on the
inertial reference frame spaces 
and from the transformation rules. However, instead of the explicit use of 
the transformation rules it is convenient to refer to simpler and more 
transparent facts expressed in the transformation rules. For instance, 
in the non-relativistic case we accept

4NR. Absolute time and absolute Euclidean structure (on absolute 
simultaneous spacetime points) exist. 

In the special relativistic case  we accept that

4SR. There exists a propagation mode of electromagnetic field
configurations, solution of Maxwell equations, and denominated
light, which is absolute (independent of the source) and is 
described by a Lorentzian structure (involving the Euclidean structure).  

Then in the non-relativistic spacetime model (NRM) and in the special
relativistic spacetime model (SRM) built up on the corresponding assumptions,
 it becomes a quasi trivial fact that 4NR and 4SR imply the Galilean 
and the Lorentzian transformation rules, respectively
(see [12], \S~I.8.2.5 and \S~II.7.1.6.).

Light phenomena are not well described in the NRM; superluminal
phenomena are not well described in the SRM. Thus if we want to treat
superluminal phenomena, we have to construct a new spacetime model
which, similarly to the known cases, will be built up on
straightforward theoretical assumptions resulting in a definite
transformation rule. Now we accept that

\smallskip
\item \llap{4W. a)} Light propagation (in the luminal model---see \S~3.2)
is absolute (independent of source),

\item  \llap{b)} There exist electromagnetic field
configurations (EFC) that can propagate at arbitrary speeds with
respect to material objects (observers),

\item  \llap{c)} there are electromagnetic field configurations which
cannot be at rest with respect to material objects (observers).

4Wb and 4Wc are suggested by the recent theoretical
investigations of [2--6] showing the existence of arbitrary
speeds ($o\leq v < \infty$) solutions of Maxwell equations.

\subhead 3. CONSTRUCTION OF A NEW SPACETIME MODEL \endsubhead 
\nopagebreak 

\subsubhead  3.1 Absolute simultaneity \endsubsubhead
\nopagebreak 

 As it is mentioned, we start with the fact that spacetime is a four 
dimensional oriented affine space $M$ (over the vector space $\M$).

The possibility of EFC of arbitrary speeds allows us to establish 
an absolute simultaneity  $S$ on $M$ by a limit procedure using EFC 
whose speed tends to infinity.  Absolute simultaneity is an equivalence relation 
 on $M$; then the set of simultaneity classes, $I:=M/S$ is {\bf absolute time}, 
 the canonical surjection $\tau:M\to I$ is {\bf time evaluation}. 

It is not a hard assumption that simultaneity classes are parallel hyperplanes, 
which implies the existence of a three dimensional linear subspace $\E$ of $\M$ 
such that  $\tau(x)=\tau(x+\E)$ for all $x\in M$. Then $\I:=\M/\E$ is a one 
dimensional vector space and $I$ becomes an affine space over $\I$ by the 
subtraction $(x+\E)-(y+\E):=x-y + \E$. Then the time evaluation $\tau$  will be an affine 
map over the canonical linear surjection $\ta:\M\to\I$. Keep in mind that $\E$ 
is the kernel of $\ta$, i.e. $\ta\cdot\boldkey x=0$ if and only if 
$\boldkey x\in\E$. 

$\E$ is the vector space of {\bf absolute spacelike} vectors; from  property 
2b in the previous paragraph we accept that there is an orientation on $\E$. 
The orientation of $\M$ and the orientation of $\E$ determine an orientation 
of $\I$ as follows. Let $(\e_0,\e_1,\e_2,\e_3)$ be a positively oriented basis 
of $\M$ such that $(\e_1,\e_2,\e_3)$ is a positively oriented basis of $\E$. 
Then $\ta\cdot\e_0$ is considered to be positive in $\I$. It is not hard to see 
 that the definition of the orientation of $\I$ does not depend on the basis. 
The orientation of $\I$ gives the orientation (an ordering) of $I$ which we 
interpret expressing future and past: $t'$ is {\it later} than $t$ if 
$t'-t>0$.

Recapitulating our results, we have

\smallskip
\item \llap{$\bullet$~}
spacetime, a four dimensional oriented affine space
$M$ (over the vector space $\M$),

\item \llap{$\bullet$~}
absolute time, a one dimensional oriented affine space $I$ (over the 
vector space $\I$),

\item \llap{$\bullet$~}
time evaluation, an affine surjection $\tau:M\to I$ (over the linear 
map $\ta:\M\to\I$), and $\E:=\text{Ker}\ta$.

\smallskip

We call attention to the following fact: in usual treatments time is
considered to be the real line but, evidently, e.g. the real number 3 is 
neither a time point nor a time period; we have got that time is an 
oriented one dimensional affine space $I$ and time periods are positive elements
of the oriented one dimensional vector space $\I$; we shall see that 
distances, too, will be positive elements of an oriented one dimensional 
vector space $\D$. Oriented one dimensional vector spaces will be called
{\bf measure lines}. We need the products and quotients of elements of
different measure lines; e.g. if $m\in\D$ and $s\in\I$, we need $m/s$.
There is a convenient mathematical expression of such products and quotients
(see Introduction of ref.~[11]) which is not detailed here because formally we can 
apply the well known rules of multiplication and division.

\subsubhead  3.2 Absolute velocities \endsubsubhead

 We have got $M$, $I$ and $\tau$ which form a part of NRM (see \S~I.1 of [11]); 
 so we can use all the notions of NRM 
that do not refer to the Euclidean structure. In particular, $r:I\to M$ 
is a {\it world line function}, if $\tau(r(t))=t$ for all $t\in I$. 
Then its derivative, the absolute velocity has the property $\ta\cdot 
\dot r(t)=1$;  correspondingly,
$$
V(1):=\left\{\u\in \frac{\M}{\I}\biggm| \ta\cdot\u=1\right\} \tag 1
$$
is the set of {\bf absolute velocities}.    
   
In contrast to the NRM, in our theory  not all world lines are allowed as
histories of mass points. According to our assumption 4W.c, the possible
particle velocities form a non void subset $P$ of $V(1)$. The
elements of $P$, 
$\partial P$ and $V(1)\setminus \overline P$ are called
{\it particle (or subluminal)} 
velocities,  {\it luminal} velocities and {\it superluminal velocities}, 
respectively.  Keep in mind that here velocity means absolute velocity.

We suppose that $P$ is open and connected. A {\bf reference
frame} is a smooth 
map $\boldkey U:M\to P$.\footnote{This definition is the one used in
[13,14,15,12]. Note that what we define as a reference frame has
been called observer in [11,12].}
Then the {\bf space of the reference frame} is as in NRM: 
it consists of the integral curves of the vector field $\boldkey U$. 
Each one of the integral lines of $\boldkey U$ is called an
{\bf observer}.
{\bf Inertial reference frames} are the ones having constant value.
In the following 
we shall deal only with inertial reference frames, so we omit the term inertial, 
and we refer to an inertial observer by its constant value,  so we say, e.g.,
an observer $\u\in P$. The $\u$-space consists of the straight lines 
parallel to $\u$; thus a $\u$-space point is of the form $x+\u\otimes\I$
for some $x\in M$.

\subsubhead  3.3 Observer times \endsubsubhead

The flow of time as registered by physical clocks are different
depending on the motion of the clocks.
(This is an experimental fact [14,17].) Consider the world 
lines of two (pointlike) clocks with velocity $\u$ and $\u_o$, respectively. 
Establish a synchronization of the clocks by an "infinitely" fast superluminal 
signal. Later the 
synchronization is repeated, and it is found that the times registered by to the 
clocks between the two synchronizations are different. Because of the 
affine structure of observer times (property 1a in \S~2) this means 
that  to every $\u\in P$ there is a positive number $\l_\u$ in such a way that 
the time elapsed between the absolute 
timepoints $t_1$ and $t_2$ along the world line with velocity value $\u$ 
equals $\frac{t_2-t_1}{\l_\u}$. 

Now we conceive that the observers in $\u$ considers time $I$ to have 
an affine structure with the $\u$-subtraction 
$$
(t_2-t_1)_\u:=\frac{t_2-t_1}{\l_\u} \tag 2
$$

\subsubhead  3.4  Observer spaces \endsubsubhead

Spaces of different reference frames are different. However, all the 
reference frames
spaces can be made an affine space over the same vector space $\E$. Take
two $\u$-lines $q_1$ and $q_2$ (representing the endpoints of a rod resting
in $\u$-space). Then 
the vector between simultaneous points of $q_2$ and $q_1$ is independent of
time. In NRM where the 
Euclidean structure is taken to be absolute, this vector is accepted to
be the difference of $q_2$ and $q_1$, defining the affine structure of 
the observer space. Now we take into account 
that the Euclidean structure depends on observers. Let us consider two
sets of observers, in two frames $\u_o$ and $\u$, both having a resting rod of the same length  $d$ 
(the number of molecules of the given crystal along the rod is the same). 
Now the observer in $\u$ marks the endpoints of the $\u_o$-rod at a given 
instant (i.e. simultaneously) and measures the distance between the marks 
and finds eventually that it does not equal $d$. Thus the two observers  
assign different vectors in $\E$ to the "same" rod.

In view of the fact 2a in paragraph 2, we assume 
that this difference can be expressed by a linear map which means the
following.

To every $\u\in P$ there is given a linear bijection $A_\u:\E\to\E$ in such 
a way that  
the observer space $E_\u$ (the set of straight lines parallel to $\u$) is 
equipped with an affine structure by the subtraction 
$$
q_2-q_1:= A_u\cdot(x_2-x_1) \qquad\qquad (x_2\in q_2,x_1\in q_1, 
\tau(x_2)=\tau(x_1)) \tag 3
$$

Since $\E$ is oriented, there is an $\E\land\E\land\E$ valued canonical 
translation invariant measure on $\E$ such that the polyhedron spanned 
by the positively oriented basis $(\e_1,\e_2,\e_3)$ equals $\e_1\land\e_2 
\land\e_3$.
$\E\land\E\land\E$ is an oriented one dimensional vector space, so we can 
take its cubic root $\D$ (see \S~IV.4. of [11]). Evidently, the elements
of $\E\land\E \land\E$ are interpreted as volume values, so the elements of
$\D$ are distances.
 
According to item 2c in paragraph 2, every observer $\u$ has a Euclidean 
structure $\bold b_\u$. If $\r$ and $\r_o$ are the same (arbitrary) rods 
in $\u$-space and $\u_o$-space, respectively, then they have the same length 
according to $\u$ and $\u_o$, respectively. Thus the Euclidean structures 
of the observer spaces define a unique Euclidean structure $\bold b:
\E\times\E\to\D\otimes\D$ such that $\bold b_\u(\r,\r) = \bold b(A_\u\cdot\r,
A_\u\cdot\r)$.

\subsubhead  3.5 Continuity \endsubsubhead

The specific meaning of the set of particle velocities $P$ in $V(1)$ 
is reflected by the fact, that we require $\u\to\l_\u$ and $\u\to A_\u$ to 
be continuous  and continuously inextensible to the points of 
$\partial P$ in such a way that they remain positive and non-degenerate, 
respectively.

\subsubhead  3.6 The new spacetime model \endsubsubhead

Recapitulating our results, we see that we have got the Euclidean 
structure on $\E$, so all the items of NRM are present, and further 
structures are introduced. We have as a new spacetime model
$$
(M,I,\tau,\D,\bold b, P, \l, A)
$$
where 

\smallskip
\item \llap{$\bullet$~}
$(M,I,\tau,\D,\bold b)$ is a NRM (in which the set $V(1)$  of absolute 
velocities is defined), and 

\item \llap{$\bullet$~}
$P\subset V(1)$ is a nonvoid connected open subset,

\item \llap{$\bullet$~}
$\l:P\to\Bbb R^+,\quad\u\mapsto\l_\u$,

\item \llap{$\bullet$~}
$A:P\to \text{GL}(\E), \quad\u\mapsto A_\u$

\smallskip

\noindent
are continuous functions which cannot be extended continuously to the points 
of $\partial P$.

\subsubhead  3.7 Notations  \endsubsubhead

The action of a linear map is denoted by a dot e.g $\ta\cdot\x$.
In the following, instead of $\bold b$ we shall write a dot
product, too, i.e.
$\q\cdot\r:=\boldkey b(\q,\r)$; furthermore, we put $|\q|^2:=
\boldkey b(\q,\q)$ for $\q\r\in\E$, and similar notations will
be applied for the induced Euclidean structure on $\frac{\E}{\I}$ etc 
(see \S~I.1.4.2 of [11]). If one treats linear maps and bilinear maps as
tensors then all these dots correspond to contractions and no ambiguity arises.

\subhead 4. SOME FORMULAE IN THE NEW SPACETIME MODEL \endsubhead 

\subsubhead  4.1. Splitting of spacetime  \endsubsubhead

Observers in the reference frame $\u\in P$ splits spacetime into time and space in such a way that
to a spacetime point $x$ they assigns the corresponding absolute timepoint
and the $\u$-spacepoint that $x$ is incident with, i.e. the $\u$-{\bf 
splitting of spacetime} is the map
$$
H_\u:M\to I\times E_\u,\qquad x\mapsto (x+\E,x+\u\otimes\I).\tag 4 $$

This splitting is the same as in NRM. However, since the affine structure 
of $E_\u$ differs
from that in NRM, and we have to consider the affine structure of $I$ 
depending on the observer (see 3.4.),now we find that $H_\u$ is an affine map
over the linear map
$$
\s_\u:\M\to\I\times\E,\qquad \x\mapsto \left(\frac{\ta\cdot\x}{\l_\u}, 
A_\u\cdot\big(\cdot \x-(\ta\cdot\x)\u\big)\right) \tag 5
$$
which we call the $\u$-{\bf splitting of vectors}.

\subsubhead  4.2 Relative  velocities \endsubsubhead

A world line function represents the history of a mass point or a light ray
signal in spacetime. An observer perceives this history as a
motion. The motion 
relative to the reference frame $\u\in P$ corresponding to the world line function
$r$ is described by the function $r_\u$ which assigns to a timepoint $t$ the 
$\u$-space point that $r(t)$ is incident with:
$$
r_\u:I\to E_u,\qquad t\mapsto r(t)+\u\otimes\I.\tag 6
$$

The velocity of the motion relative to the observer is obtained by
$$
\lim_{t_2\to t_1}\frac{r_\u(t_2)-r_\u(t_1)}{(t_2-t_1)_\u}=
\l_\u A_\u\cdot(\dot r(t_1)-\u).  \tag 7
$$

That is why we accept that if $\w\in V(1)$ and $\u\in P$ then
$$
\v_{\w\u}:=\l_\u A_u\cdot(\w-\u) \tag 8
$$
is the {\bf relative velocity of $\w$ with respect to $\u$}.

Then we have for $\u,\u'\in P$
$$
\l_\u'A_\u^{-1}\cdot\v_{\u'\u}=-\l_\u A_{\u'}^{-1}\cdot\v_{\u\u'} \tag 9
$$
which implies, in general, that $\v_{\u\u'}\neq-\v_{\u'\u}$.

Furthermore, we easily find the {\bf velocity addition formula}: 
if $\w\in V(1)$ and $\u,\u'\in P$ then
$$
\v_{\w\u}=\v_{\u'\u} + \frac{\l_\u}{\l_{\u'}}A_\u\cdot A_{\u'}^{-1}\cdot
\v_{\w\u'} \tag 10
$$
or
$$
\frac{\l_{\u'}}{\l_\u}A_{\u'}\cdot A_\u^{-1}\cdot\v_{\w\u}=
\v_{\w\u'}-\v_{\u\u'}.  \tag 11
$$

\subsubhead  4.3 Comparison of splittings \endsubsubhead

Let us compare now the splittings in two different reference
frames $\u,\u'\in P$
which is expressed by $\s_{\u'}\cdot\s_\u^{-1}$. Since $\s_\u^{-1}(\t,\q)=
A_\u^{-1}\cdot\q + \frac{\t}{\l_\u}\u$, we easily find the {\bf vector 
transformation law}:
$$
\s_{\u'}\cdot\s_u^{-1}(\t,\q)=\left(\frac{\l_\u}{\l_{\u'}}\t,
A_{\u'}\cdot A_\u^{-1}\cdot\q + \v_{\u\u'}\frac{\l_\u}{\l_\u'}\t\right);
 \qquad (\t,\q)\in\I\times\E. \tag 12
$$

We call the reader's attention to the fact that here $\v_{\u\u'}$ cannot
be substituted with $-\v_{\u'\u}$; if we want to use the latter quantity,
we obtain
$A_{\u'}\cdot A_\u^{-1}\cdot(\q - \v_{\u'\u}\t)$ in the formula of the 
transformation law.

\subhead 5. THE LORENTZ AETHER MODEL (LAM) [18,19] \endsubhead 

\subsubhead  5.1 Aether, dilation, contraction \endsubsubhead

 The previous type of spacetime model is very general (it contains 
the NRM as a particular case: then $P=V(1)$, $A_\u$ is the identity of $\E$ 
and $\l_\u=1$ for all $\u$). Now we shall detail a special model (LAM) where

\smallskip
\item \llap{$\bullet$~}
there are an $\u_o\in P$  and a $0<c\in \frac{\D}{\I}$ such that
$$
P = \left\{\u\in\frac{\M}{\I}\biggm| |\v_{\u\u_o}|<c\right\} =
\u_o +\left\{\v\in\frac{\E}{\I}\biggm| |\v|<c \right\}, \tag 13
$$
$$
\l_\u=\frac1{\sqrt{1-\frac{|\v_{\u\u_o}|^2}{c^2}}} \tag 14
$$
$$
A_\u=\1 - (1-\l_\u) \frac{\v_{\u\u_o}}{|\v_{\u\u_o}|}\otimes
\frac{\v_{\u\u_o}}{|\v_{\u\u_o}|}  \tag 15
$$

Regarding the previous definition, note that the symbol $\1$ denotes the
identity map of $\E$ and for $\boldkey n$ the linear map
$\boldkey n\otimes\boldkey n$ acts as $\q\mapsto\boldkey n(\boldkey n\cdot\q)$.

We find that $\l_{\u_o}=1$; furthermore
if $\u=\u_o$ then the expression containing  $|\v_{\u\u_o}|=0$ in the 
denominator is meaningless but it is multiplied 
by zero, so we mean that $A_{\u_o}=\1$.

Of course, the set of luminal velocities is 
$$
\partial P= \left\{\w\in\frac{\M}{\I}\biggm| |\v_{\w\u_o}|=c\right\}. \tag 16
$$

The reference frame with constant velocity $\u_o$ is called the {\bf aether},
$c$ is the {\bf light speed in the aether}, $\l_\u$ is the {\bf time
dilation factor} corresponding to $\u$, and 
$$
A_\u^{-1}=\1 +\frac{1-\l_\u}{\l_\u} \frac{\v_{\u\u_o}}{|\v_{\u\u_o}|}
\otimes\frac{\v_{\u\u_o}}{|\v_{\u\u_o}|}   \tag 17
$$ 
is the {\bf Lorentz contraction map} corresponding to $\u$: $|A_\u^{-1}\cdot\q|
=|\q|$ if $\q$ is
orthogonal to $\v_{\u\u_o}$ and $|A_\u^{-1}\cdot\q|=\l_\u|\q|$ if $\q$ is 
parallel to $\v_{\u\u_o}$. 

\subsubhead  5.2 Relative velocities \endsubsubhead

The equality 
$$
\v_{\w\u_o}=\w-\u_o  \tag 18
$$
is a trivial fact for $\w\in V(1)$; in general, if $\u\in P$ then
$$
\v_{\w\u}=\l_\u A_u\cdot(\v_{\w\u_o}-\v_{\u\u_o}).   \tag 19
$$

Having the LAM, we can calculate quite easily
all the quantities appearing in usual applications of aether theory
[18--26] without further assumptions and heuristic considerations.
For instance, we have for $\w\in V(1)$, $\u\in P$
$$
|\v_{\w\u}|^2=\l_\u^2\left[|\v_{\w\u_o}|^2 + \l_\u^2\left(|\v_{\u\u_o}|^2 
-2\v_{\w\u_o}\cdot\v_{\u\u_o} +
\frac{(\v_{\w\u_o}\cdot\v_{\u\u_o})^2}{c^2}\right)\right].  \tag 20
$$

We see that for $\u,\u'\in P$
$$
|\v_{\u'\u}|\neq|\v_{\u\u'}|\qquad\text{in general},  \tag 21
$$
more closely,
$$
|\v_{\u'\u}|=|\v_{\u\u'}|\qquad\text{if and only if}\qquad
|\v_{\u'\u_o}|=|\v_{\u\u_o}|.  \tag 22
$$

In particular, we have
$$
|\v_{\u_o\u}|=\l_u^2|\v_{\u\u_o}|.  \tag 23
$$

\subsubhead  5.3 Ives-Tangherlini-Marinov coordinates [16,18,19,21,22,23] \endsubsubhead

If we choose a positively oriented basis $(\e_0,\e_1,\e_2,\e_3)$ in $\M$
such that $(\e_1,\e_2,\e_3)$ is a positively oriented orthogonal basis in $\E$,
$\e_0$ is parallel to $\u_o$, $\e_1$ is parallel to $-\v_{\u_o\u}$ (which
is not equal to $\v_{\u\u_o}$!) then
the transformation law given in 4.1 applied to $\u':=\u_o$ and expressed
in coordinates relative to the chosen basis coincides with the well known 
Ives-Tangherlini-Marinov transformation.

\subhead 6. THE RELATIVISTIC STRUCTURE DUE TO THE AETHER \endsubhead 

\subsubhead  6.1 The Lorentz form \endsubsubhead

Due to the privileged observer (aether) in the LAM we can introduce a 
Lorentz form on $\M$ by the use of $\u_o$-splitting:
$$
\x\cdot\y:=(\x-(\ta\cdot\x)\u_o)\cdot(\y-(\ta\cdot\y)\u_o) -
c^2(\ta\cdot\x)(\ta\cdot\y).   \tag 24
$$

The Lorentz product denoted by a dot on the left hand side is an extension 
of the Euclidean dot product appearing on the right hand side, so the
notation is consistent.

The Lorentz form is arrow oriented in such a way that $\u_o$ is
pointing to the future.

So $(M,\D,\cdot)$ is a SRM {\it associated} to the LAM
in which all the well known relativistic notions can be used
([11], Part II).

\subsubhead  6.2 Relativistic splitting \endsubsubhead

For $\w,\w'\in V(1)$, we have
$$
-\w'\cdot\w=c^2-\v_{\w'\u_o}\cdot\v_{\w\u_o}.   \tag 25
$$

In particular, $-\u_o\cdot\w=c^2$ for all $\w\in V(1)$. Moreover, it
follows that 
$$
\left\{\hat\u\in \frac{\M}{\D}\biggm| \hat\u\text{ is future directed },
-\hat\u\cdot\hat\u<1\right\}=\left\{\frac{\l_\u\u}{c}\biggm| \u\in P\right\}.
\tag 26
$$

As a consequence, the inertial reference frames of the LAM
coincide with the inertial reference frames of the associated SRM. For the sake 
of brevity, we introduce the notation
$$
\hat\u:= \frac{\l_u\u}{c} \qquad\qquad (\u\in P).   \tag 27
$$

We mention that $\hat\u_o=-\frac{\u_o}{c}$ and 
$$
\ta\cdot\x=-\frac{\u_o\cdot\x}{c^2}\qquad\qquad (\x\in\M).  \tag 28
$$

The relativistic synchronization established by luminal phenomena 
 depends on the reference frame. The reference frame 
$\u\in P$ finds that luminally simultaneous spacetime points are
hyperplanes parallel to the three dimensional subspace
$$
\E_\u:=\{\x\in\mid \u\cdot\x=0\}.  \tag 29
$$

Using this synchronization, the space of the reference frame $\u$
(the set of 
straight lines parallel to $\u$) becomes an affine space over $\E_\u$
by the subtraction 
$$
(q_2-q_1)_{\text rel}:=x_2-x_1 \qquad (x_2\in q_2,x_1\in q_1, \u\cdot(x_2-x_1)
=0).   \tag 30
$$

Thus $\u$-space vectors are different in the LAM and in the
associated SRM. This important fact disappears
when considering coordinates, since coordinates of arbitrary three 
dimensional vector spaces are triplets of numbers.

The set $I_\u$ of hyperplanes parallel to $\E_\u$ constitute
the time of the reference frame; this is a one dimensional affine
space over $\I$ by the subtraction
$$
 t_2-t_1:=-\frac{\hat\u}{c}\cdot(x_2-x_1) \qquad\qquad (x_2\in t_2,
 x_1\in t_1).\tag 31
$$

According to the relativistic synchronization, the observers in
the reference frame $\u$ split spacetime in time and space by
$$
	H_{\hat\u}: M \to I_{\u} \times E_{\u}, \qquad
		x \mapsto (x+E_{\u},x+\u\otimes I)	\tag 32
$$
which is an affine mapping over the linear map
$$
\h_{\u}:\M\to\I\times\E_\u,\qquad \x\mapsto \left(\frac{-\hat u\cdot\x}{c},
\x+(\hat u\cdot\x)\hat u\right).  \tag 33
$$

As an important fact, we mention that the relativistic relative velocity 
of $\u'\in P$ with respect to $\u\in P$ is (see [11], \S~II.4.2.)
$$
\v_{\hat\u'\hat\u}:=c\left(\frac{\hat\u'}{-\hat\u\cdot\u'}-\hat\u\right).
\tag 34
$$

\subsubhead  6.3 Lorentz boosts \endsubsubhead

The relativistic spaces of different reference frames $\u$ and $\u'$
are affine spaces over the different vector spaces $\E_\u$ and $\E_{\u'}$,
respectively. However, there is a ``canonical'' linear bijection
between $\E_\u$ and $\E_{\u'}$ which can be used to identify these different
vector spaces; this linear bijection is the {\bf Lorentz boost} from 
$\u$ to $\u'$ (see \S~II.1.3.8 of [12])
$$
L(\u',\u):=\1 + \frac{(\hat\u'+\hat\u)\otimes(\hat\u'+\hat\u)}
{1-\hat\u'\cdot\hat\u} -2\hat\u'\otimes\hat\u    \tag 35
$$
where $\1$ is the identity map of $\M$.

This linear bijection preserves the Lorentz form and its arrow orientation
as well as the orientation of spacetime. Moreover, we have
$$
\eqalign{
& L(\u',\u)\cdot\hat\u=\hat\u'  \cr
& L(\u',\u)\cdot\q=\q \quad \text{if}\ \q\in\E_\u\cap\E_{\u'} \cr
& L(\u',\u)\cdot\v_{\hat\u'\hat\u}=-\v_{\hat\u\hat\u'} \cr
& L(\u',\u)^{-1}=L(\u,\u') \cr }   \tag 36
$$
In usual treatments based on coordinates the space of every reference
frame is considered to consist of the elements of
form $(0,\xi_1,\xi^2,\xi^3)$.
This corresponds to the fact that one chooses a reference frame
(``rest frame'') and implicitly all the other reference frames spaces
are Lorentz boosted to the space of that reference frame. 

\subsubhead  6.4 Comparison of splittings \endsubsubhead

If superluminal phenomena do exist then the Lorentz aether model offers
an adequate structure for spacetime. Then we conceive that the relativistic
formulae used in physics refer to the SRM associated to the LAM. Therefore
it is important to compare the splitting $\s_\u$ in LAM and 
the splitting $\h_\u$ in SRM due to a reference frame $\u\in P$.
Since $\h_\u^{-1}(\t,\q)=\q + \hat\u c\t$ for $\t\in\I$, $\q\in\E_\u$,
 we easily find the comparison:
$$
\s_\u\cdot\h_\u^{-1}(\t,\q)=\left(\t+\frac{\ta\cdot\q}{\l_\u},
\q - (\ta\cdot\q)\u\right).   \tag 37
$$

However, this is rarely useful, because relates elements in $\E_\u$ to
elements in $\E=\E_{\u_o}$. To have a nicer formula, we map $\E_\u$ onto
$\E_{\u_o}$ "canonically", i.e. we apply a Lorentz
boost from $\u$ to $\u_o$, and instead of the splitting $\h_\u$ we
consider 
$$
\h_{\u_o\u}:=\biggl(\text{id}_{\I}\times
L(\u_o,\u)|_{\E_{\u}}\biggr)\cdot\h_\u= 
\h_{\u_o}\cdot L(\u_o,\u)   \tag 38
$$
(see \S~II.7.1.4--7.1.6 of [11]) for which
$$
\h_{\u_o\u}\cdot\x=\left(-\frac{\hat\u}{c}\cdot\x,
L(\u_o,\u)\cdot\x + (\hat\u\cdot\x)\hat\u\right)\tag 39
$$
holds, and we look for the explicit expression of
$$
\s_\u\cdot\h_{\u_o\u}^{-1}=\s_\u\cdot L(\u,\u_o)\cdot\h_{\u_o}^{-1}  \tag 40
$$
applied to elements $(\t,\q)\in \I\times \E_{\u_o}$.

The time component, by $L(\u.\u_o)\cdot\h_{\u_o}^{-1}(\t,\q)=
L(\u,\u_o)\cdot(\q+\u_o\t)=L(\u,\u_o)\cdot\q + \l_u\u\t$, by
Eqs.~5, ~27 and ~34 and by the fact that $\u\cdot
L(\u,\u_o)\cdot\q_o=0$, is obtained as
$$
\t+\frac{\v_{\u\u_o}}{c^2}\cdot\q.   \tag 41
$$

Furthermore we obtain by simple calculations that
$$
A_{\u}=L(\u_o,\u)(\1 + \hat\u\otimes\hat\u)|_{\E_{\u_o}}   \tag 42
$$
and taking into account the formulae
$$
\x-(\ta\cdot\x)\u= \left(\1 + \frac{\hat\u\otimes\hat\u_o}{c\l_\u}\right)
\cdot\x,   \tag 43
$$
$$
(\1+\hat\u\otimes\hat\u)\cdot\left(\1+\frac{\hat\u\otimes\hat\u_o}{c\l_\u}
\right)=\1+\hat\u\otimes\hat\u,   \tag 44
$$
we find that the space component equals $\q$; summarizing our results:
$$
\s_{\u}\cdot\h_{\u_o\u}^{-1}(\t,\q)=\left(\t+\frac{\v_{\u\u_o}}{c^2}\cdot\q,
\,\, \q\right)  \qquad \bigl((\t,\q)\in \I\times \E_{\u_o}\bigr).  \tag 45
$$

\subsubhead  6.5 Comparison of motions in LAM and in SRM \endsubsubhead

The history of a masspoint given by a world line function
$r:I\mapsto M$ is perceived by an observer in $\u$ as a motion;
the motion is described in different ways in LAM and in SRM. To get a
better comparison between the different descriptions, we
consider a vectorization of spacetime by an origin $o$, i.e. 
the vectorized motion $\r_\u$ in LAM is obtained from
$$
	\s_\u(r(t)-o) = (t-t_o, r(t)-o+\u(\t-t_o))   \tag 46
$$
where $t_o:=\tau(o)$; thus by $\t:=t-t_o$ we get 
$$
	\r_\u:\I\to \E,\qquad \t\mapsto r(t_o+\t)-o+\u\t  \tag 47
$$

The vectorized motion $\r_{\hat\u}$ in SRM (applying a boost to $\u_o$) is
obtained from
$$
h_{\u_o\u}(r(t)-o)= \biggl(-\frac{\hat\u}{c}\cdot(r(t)-o),\,\,
L(\u_o,\u)\cdot(r(t)-o) + (\hat\u\cdot(r(t)-o))\hat\u\biggr). \tag 48
$$
Since
$$
	\t_\u:=-\frac{\hat\u}{c}\cdot(r(t)-o)\tag 49
$$
gives the relativistic $\u$-time as a function of absolute time
$t$ from which we can express  $t$ as a function of $\t_\u$, we
get the vectorized motion in SRM: 
$$
	\r_{\hat\u}:\I\to \E,\qquad
		\t_\u\mapsto L(\u_o,\u)\cdot(r(t(\t_u))-o)
			+ (\hat\u\cdot(r(t(t_\u)-o))\hat\u. \tag 50
$$

Then these formulae or the one at the end of the previous
paragraph allow us to recover the LAM motion from the SRM
motion: the function
$\t\mapsto\t+\frac{\v_{\u\u_o}}{c^2}\cdot\r_{\hat\u}\t)$ is continuously 
differentiable, its derivative is everywhere positive, so it
has a continuously differentiable inverse, denoted by
$\s\mapsto\t(\s)$, and we have
$$
	\r_\u(\s)=\r_{\hat\u}(\t(\s))   \tag 51
$$
which implies
$$
\dot\r_\u(\s)=\frac{\dot\r_{\hat\u}(\t(\s))}{1+(\v_{\u\u_o}/c^2)\cdot
\dot\r_{\hat\u}(\t(\s))}.   \tag 52
$$

\subsubhead  6.6 Propagation of superluminal waves \endsubsubhead

The Lorentz invariance of the Maxwell equations means in our language
that the relativistic split form of the absolute Maxwell equations is
the same for all observers. Thus time, space and velocity in a solution
of the split Maxwell equations concern the relativistic splitting due to
an observer $\u$. Now we want to express the solution in quantities
corresponding to the aether splitting. Since the usual form of the solutions
is given in coordinates which means that all the quantities are automatically
 boosted to
a "basic" observer, $\u_o$ in our notations, the result of the previous 
paragraph says us that passing from the relativistic splitting (coordinates)
to the aether splitting, space vectors remain unchanged and the relativistic
time $\t$ is to be substituted with $\t-\frac{\v_{\u\u_o}}{c^2}\cdot\q$. 

Suppose  now that we are given a solution of the Maxwell equations relative 
to the observer $\u$, and the solution describes a wave propagating with
velocity $\v$. The wave propagation
The wave propagation corresponds a uniform
motion with velocity $\v$ in SRM, thus we infer from the result
at the end of the previous paragraph that the relative velocity
in LAM equals
$$
\frac{\v}{1+\frac{\v_{\u\u_o}\cdot\v}{c^2}}.\tag 53
$$

The denominator must be positive, which means that for an observer $\u\neq\u_o$ 
not all elements of $\frac{\E_{\u}}{\I}$ are allowed as relativistic 
relative velocities.
Regarding in the reversed order, we can say that all elements of
$\frac{\E_{u_o}}{\I}$  can be relative velocities with respect to
an arbitrary observer $\u$ in the LAM but their 
transforms in the associated SRM do not fill the whole $\frac{\E_\u}{\I}$.

\subsubhead  6.7 An application to rotating bodies \endsubsubhead

There is a long dispute on whether the Lorentz aether theory or special
relativity is the adequate theory of spacetime. If superluminal phenomena
will be detected then there is no doubt. If not, the present mathematical 
model may help us to answer the question ruling out loosely
defined notions and tacit assumptions regarding Lorentz aether theory
which can be found in most of the reasonings (e.g., in [24,25]) as it is
pointed out in [19].

The experiments proposed in [24,25] refer to uniformly rotating rigid bodies.
However, as it turns out (see \S~II.6.7--6.8 of [12]), the relativistic theory
does not admit an object which would have all the well known usual properties
of a nonrelativistic uniformly rotating rigid body, so we must be very
cautious in reasonings regarding them.

Let $o$ be a spacetime point, $B_o$ be a subset of $\E=\E_{\u_o}$ and 
$\Omega_o:\E\to\E$ 
 an antisymmetric linear map. Then the collection of world lines
$$
\{\t\mapsto o + \l_\u\u\t + A_\u\cdot\exp(\t\Omega_o)\cdot\q_o\mid
\q_o\in B_o\}  \tag 54
$$ 
gives a uniformly rotating rigid body in the space of the reference
frame $\u$ 
according to the LAM.

It seems, the "uniformly rotating reference frame II" described
in [12], \S~II.6.8.  is the best candidate to be accepted as a
uniformly rotating relativistic rigid body. This describes an
object which is seen uniformly rotating relative to $\u\in P$.
All its points are given by a world line of the form
$$
\t\mapsto r(\t):= o+ \hat\u\t + \exp(\t\Omega)\cdot\q   \tag 55
$$
where $o$ is a given spacetime point, $\Omega$ is an antisymmetric linear
map $\E_\u\to\frac{\E_\u}{\I}$ and $\q\in\E_\u$ is in the kernel of
$\Omega$,  $\t$ is the (relativistic) time of the reference
frame $\u$ passed 
from the $\u$-timepoint corresponding to $o$; lastly, the inequality
$\omega|\q|<c$ must be satisfied where $\omega:=|\Omega|$.

The $\u$-splittings of $\t\mapsto r(\t)-o$ in SRM and in LAM give the
corresponding motion relative to $\u$.

The relativistic motion (see \S~6.5) is indeed a uniform rotation
$$
	\t\mapsto\exp(\t\Omega_o)\cdot\q_o	\tag 56
$$
where $\Omega_o:=L(\u,\u_o)\cdot\Omega\cdot L(\u,\u_o)$ and
$\q_o:=L(\u,\u_o)\cdot \q$.

As concerns the motion in LAM, we have to find the inverse of
the function 
$$
	\t \mapsto s_\q(\t) := \t+\frac{\v_{\u\u_o}}{c^2} \cdot
				\exp(\t\Omega_o) \cdot \q_o \tag 57
$$

By a convenient choice of the "origin" $o$ we can attain that $\v_{\u\u_o}$
be orthogonal to $\Omega_o\cdot\q_o$; thus, since $\exp(\t\Omega_o)\cdot\q_o=
\q_o\cos\omega\t + \frac{\Omega_o\cdot\q_o}{\omega}\sin\omega\t$,
we find that
$$
	\s_\q(\t)=\t + \frac{\v_{\u\u_o}}{c^2}\cdot\q_o\cos\omega\t.  \tag 58
$$

If we denote the inverse of $\s_{\q}$ by $\s\mapsto t_\q(\s)$, then the motion
relative to the reference frame becomes, according to the LAM
$$
\s\mapsto\exp(\t_\q(\s)\Omega_o)\cdot\q_o.  \tag 59
$$

This is not a uniform rotation. Moreover, if we take a subset $B_o$ of 
$\E$ in which
$\q_o$ can vary, the corresponding world lines form a body which
is rigid in the SRM but it is not rigid in the LAM. 

Thus in the usual considerations of uniformly rotating rigid 
bodies one should specify from what point of view the body is rigid and 
uniformly rotating. This is important in view of the rotor Doppler shift
experiments, like the Kolen-Torr experiments [24,25]. See a detailed
discussion in~[19,26].

We end this paper remarking that Santilli's isominkowskian spacetime [27-32]
provides an alternative representation of causal events with arbitrary
speeds and more particularly that his isopoincar\'e symmetry [28] provides
the explicit form of the symmetry transformations valid for arbitrary
causal speeds. The evidently expected connection between our studies and
Santilli's isominkowskian spacetime will be studied at some later time. 

%\Refs

\subhead REFERENCES \endsubhead 

\item {[1]} J. Y. Lu and J. F. Greenleaf, {\it IEEE Transact.
Ultrason. Ferroelec. 
Freq. Contr.} {\bf 39}, 19 (1992).

\item {[2]} R. Donelly and R. Ziolkowski, {\it Proc. R. Soc. London\/}
{\bf A460}, 541 (1993).

\item {[3]} A. O. Barut and H. C. Chandola, {\it Phys. Lett.}
{\bf A180}, 5 (1993). 

\item {[4]} W. A. Rodrigues, Jr. and J. Vaz, Jr., 
% ``Subluminal and superluminal solutions in vacuum of the 
% Maxwell equations and the massless Dirac equation,'' 
% RP 44/95 IMECC--UNICAMP, in publication 
{\it Adv. Appl. Clifford Algebras\/} {\bf 7} (Supl.), 453 (1997). 

\item {[5]} W. A. Rodrigues, Jr. and J. E. Maiorino, 
{\it Random Operators
and Stochastic Equations\/} {\bf 4}, 355 (1996).

\item {[6]} W. A. Rodrigues, Jr. and J. Y. Lu, 
% ``On the existence of undistorted progressive waves (UPWs) of 
% arbitrary speeds $0\leq  v  < \infty$ in nature,'' 
% RP 12/96 IMECC--UNICAMP, in publication in 
{\it Found. Phys.\/} {\bf 27}, 435 (1997).

\item {[7]} J. Y. Lu and W. A. Rodrigues, Jr., ``What is the
speed of a sound wave in a homogeneous medium?'' 
RP 25/96 IMECC--UNICAMP. 

\item {[8]} C. I. Mocanu, {\it Found. Phys. Lett.\/} {\bf 5}, 444 (1992).

\item {[9]} J. Rembieli\'nski, {\it Phys. Lett.} {\bf A78}, 33 (1980).

\item {[10]} J. Rembieli\'nski, ``Tachyon neutrino?'' preprint KFT 5/94,
Univ. Lodz, Poland, subm. for publication.

\item {[11]} T. Matolcsi, {\it A Concept of Mathematical Physics,
Models for Spacetime}, Aka\-d\'e\-miai Kiad\'o, Budapest (1984).

\item {[12]} T. Matolcsi, {\it Spacetime without Reference Frames},
Aka\-d\'e\-miai Kiad\'o, Budapest (1993).

\item {[13]} W.~A.~Rodrigues. Jr, Q.~A.~G. de Souza, and Y. Bozhkov,
{\it Found. Phys.} {\bf 25}, 871 (1995).

\item {[14]} W. A. Rodrigues, Jr. and M. A. Faria Rosa, {\it Found. Phys.}
{\bf 19}, 705 (1989).

\item {[15]} P. G. Appleby and N. Kadianakis, {\it Arch. Rat. Mech. Anal.}
{\bf 95}, 1 (1926).

\item {[16]} W. A. Rodrigues, M. Scanavini, and L. P. de Alc\^antara,
{\it Found. Phys. Lett.} {\bf 3}, 59, (1990).

\item {[17]} W. A. Rodrigues and E. C. de Oliveira, {\it Phys. Lett.}
{\bf A14}, 474 (1989).

\item {[18]} ``The Einstein Myth and the Ives Papers,''
ed. R. Hazalett and D.Turner,
The Davin-Adair Co. Old Greenwich, 1979.

\item {[19]} W.~A. Rodrigues, Jr. and J. Tiomno,
Found. Phys. 15 (1985) 845-961.

\item {[20]} T. Chang, J. Phys. A 13 (1980) 1089.

\item {[21]} F. R. Tangherlini, {\it N. Cimento Supl.} {\bf 20}, 1 (1961).

\item {[22]} S. Marinov, {\it Czech. J. Phys.} {\bf B24}, 965 (1974).

\item {[23]} S. Marinov, {\it Gen. Rel. Grav.} {\bf 12}, 57 (1980).

\item {[24]} D. G. Torr and P. Kolen, {\it Found. Phys.} {\bf 12}, 265 (1982).

\item {[25]} P. Kolen and D. G. Torr, {\it Found. Phys.} {\bf 12}, 407 (1982).

\item {[26]} A. K. A. Maciel and J. Tiomno, {\it Phys. Rev. Lett.} {\bf 55},
143 (1985).

\item {[27]} R. M. Santilli, 
{\it Lett. N. Cimento\/} {\bf 37}, 545--555 (1983).

\item {[28]} R. M. Santilli, 
{\it J.  Moscow Phys. Soc.} {\bf 3}, 255--280 (1993).

\item {[29]} R. M. Santilli, 
{\it Chinese J. of Syst. Eng. and Electr.} {\bf 6}, 157--176 (1995).

\item {[30]} R. M. Santilli, in T. L. Gill (ed.), {\it New
Frontiers in Hadronic Mechanics}, Hadronic Press (1996). 

\item {[31]} R. M. Santilli, {\it Elements of Hadronic
Mechanics}, vols. I and II (second ed.), Naukora Dumka Publ.,
Ukraine Acad. Sci., Kiev (1995).

\item {[32]} R. M. Santilli, ``Isospecial relativity with
applications to quantum gravity, antigravity and cosmology,''
{\it Balcan Geom. Press}, Budapest (in press).

%\endRefs
\bye